\documentclass[aps,prl,amsfonts,amsmath,twocolumn]{revtex4}

\usepackage{color}
\usepackage{graphicx}

\usepackage{epstopdf}

\newcommand {\be} {\begin{equation}}
\newcommand {\ba} {\begin{eqnarray}}
\newcommand {\ee} {\end{equation}}
\newcommand {\ea} {\end{eqnarray}}

\begin{document}

\preprint{WM-07-105}

\title{Hadronic Momentum Densities in the Transverse Plane}

\author{Zainul Abidin and Carl E.\ Carlson}
\affiliation{
Department of Physics, College of William and Mary, Williamsburg, VA 23187-8795, USA}

\date{\today}

\begin{abstract}
We examine the spatial density within extended objects of the momentum component $p^+$, and find relativistically exact connections to Fourier transforms of gravitational form factors.  We apply these results to obtain semi-empirical momentum density distributions within nucleons, and similar distributions for spin-1 objects based on theoretical results from the AdS/QCD correspondence.  We find that the momentum density in the transverse plane is more compact than the charge density.
\end{abstract}

\maketitle

%
%

Mapping the distribution of matter inside the nucleon is a goal of hadron structure physics.  For example, one can relate the density of charged and magnetically polarized material inside a nucleon to Fourier transforms of charge and magnetic form factors.  This old area has seen interesting and notable improvement lately, by way of finding relativistically exact relations between form factor data and two-dimensional line-of-sight projections of the charge density and polarization density~\cite{Burkardt:2002hr,Miller:2007uy,Carlson:2007xd}.

In a related fashion, as we study here, one can map out the distribution of momentum within a hadron.  Here too there is old information, in particular that (at some resolution scale) the momentum in a hadron is carried in about equal measures by quarks and gluons. More precisely, one reports the fraction $x$ of the light-front longitudinal momentum,
$p^+ \equiv p^0 + p^3$,  that is carried by individual constituents.  In an extended object, one can further ask how the total momentum is distributed point by point, either in three-dimensions, or as we shall do here, in a two-dimensional line-of-sight projection.

We will study the spatial distribution of the momentum component $p^+$,  finding relativistically exact connections to Fourier transforms of gravitational form factors.   Gravitational form factors are directly related to empirical data, in that they are matrix elements of the stress (or energy-momentum) tensor which can be obtained as second Mellin moments of generalized parton distributions (GPDs)~\cite{Ji:1996ek,Abidin:2008ku}.  The GPDs are accessible from deeply virtual Compton scattering experiments and can be loosely defined as the amplitude for replacing a parton in a hadron with one of a different light-front momentum. In the forward limit, GPDs reduce to ordinary parton distribution functions, and the first Mellin moments give the electromagnetic form factors; for reviews, see {\it e.g.}~\cite{Ji:1998pc,Goeke:2001tz,Diehl:2003ny,Belitsky:2005qn,Boffi:2007yc,Polyakov:2002yz}.

In a light-front formalism, one can kinematically connect the wave functions of any hadrons of relevant momentum. Conventionally, the longitudinal direction is the $z$-direction, chosen to point along the vector direction of $p=(p_1+p_2)/2$, where $p_1(p_2)$ is the momentum of the initial(final) hadron. Further,  one chooses the  frame 
so that the momentum transfer $q = p_2-p_1$ has $q^+=0$, and satisfies $q^2=-\vec q_\perp^2=-Q^2$. Within the light-front formalism, one can project the probability density into transverse plane~\cite{Burkardt:2000za} and show that it is obtained from the 2-dimensional Fourier transform of a form factor.

Transverse charge densities have been calculated for nucleons~\cite{Burkardt:2002hr,Miller:2007uy,Carlson:2007xd} and deuterons~\cite{Carlson:2008zc}, using empirical electromagnetic form factors. Similar to the charge density, one can observe that the $T^{++}$ component of the stress or energy-momentum tensor is the density corresponding to the $p^+$ component of the $4$-momentum,
\be
P^+ = \int T^{++} \, d^2x_\perp dx_+ \,.
\ee

In the transverse plane in position space, the density of momentum $p^+$ can be defined from expectation value of the local operator $T^{++}(x^+,x^-,x_\perp)$,
\be
\rho^+(\vec{b}_\perp)=\frac{1}{2{p^+}^2}  \left<\Psi|T^{++}(0,0,\vec{b}_\perp)\big|\Psi\right>,
\ee
such that $\int d^2 \vec{b}_\perp \rho^+(\vec{b}_\perp)=1$. The state $|\Psi\big>$ is localized with its transverse center of momentum located at the origin, and is formed by linear superposition of light-front helicity eigenstates $\big| p^+,\mathbf{p}_\perp,\lambda\big>$,
\be
|\Psi\big> \equiv |p^+,\vec{R}_\perp=0,\lambda\big>=\mathcal{N}\int \frac{d^2 \vec{p}_\perp}{(2\pi)^2}\big| p^+,\vec{p}_\perp,\lambda\big>.
\ee
Normalization factor $\mathcal{N}$ satisfies $|\mathcal{N}|^2 \int{d^2 \vec{p}_\perp}/{(2\pi)^2}=1$.

The matrix elements of the stress tensor between momentum eigenstates can be written as
\ba
\left<p^+, \frac{\vec{q}_\perp}{2},\lambda_2\bigg| T^{++}(0) \bigg|p^+,-\frac{\vec{q}_\perp}{2},\lambda_1\right>&\!=\!&
2(p^+)^2  \,  e^{i(\lambda_1-\lambda_2)\phi_q}\nonumber\\
&&\times \ \mathcal{T}^+_{\lambda_2\lambda_1}(Q^2),\label{Tmatrixelement}
\ea
where $\lambda_1(\lambda_2)$ denotes the initial (final) light-front helicity, and $\vec{q}_\perp=Q(\rm{cos}\phi_q \hat{e}_x+ \rm{sin}\phi_q \hat{e}_y)$. The $p^+$ momentum density can be written in terms of 2-dimensional Fourier transform of the form factor
\ba
\rho^+_\lambda(\vec{b}_\perp) &=& \int \frac{d^2\vec{q}_\perp}{(2\pi)^2} e^{-i\vec{q}_\perp\cdot\vec{b}}\mathcal{T}^+_{\lambda\lambda}(Q^2).
\ea

We will calculate the spatial distribution of $p^+$ in the transverse plane for spin-$1/2$ and spin-$1$ hadrons, specifically the nucleons and the rho mesons. 

{\textit {The momentum density results for nucleons}} will be semi-empirical.  We use matrix elements of the stress tensor  obtained from second moments of GPDs~\cite{Ji:1996ek}.   ``Semi'' in semi-empirical is a reminder that the GPDs are not yet measured in detail.  However, the models are constrained to accurately represent measured parton distribution functions in one limit and to give measured electromagnetic form factors in another.

For spin-1/2 particles,  matrix elements of the stress tensor involve three gravitational form factors,
\begin{align}
& \left\langle p_2, \lambda_2 \right|  T^{\mu\nu} (0)
	\left | p_1, \lambda_1 \right\rangle =
						 \bar u_{\lambda_2} (p_2)
\bigg( A(Q^2) \gamma^{(\mu}p^{\nu)} 
									\\	\nonumber
& \quad + \	B(Q^2) \frac{p^{(\mu}   i\sigma^{\nu) \alpha} q_\alpha}{2m} +
 \ C(Q^2)
\frac{q^\mu q^\nu - q^2 g^{\mu\nu}} {m}     \bigg)
u_{\lambda_1}(p_1)	,
\end{align}
where $\gamma^{(\mu}p^{\nu)} = (\gamma^\mu p^\nu + p^\nu \gamma^\mu )/2$ and $p_{1,2} = p \mp q/2$.

For the vector GPDs for spin-1/2 particles, the quarkic contribution is defined by
\begin{align}
& p^+ \int \frac{dy^-}{2\pi} e^{ixp^+y^-}
\langle p_2, \lambda_2 | \bar q(-\frac{y}{2})
\gamma^\mu  q(\frac{y}{2}) | p_1, \lambda_1
	\rangle_{y^+,y_\perp=0} =
\nonumber \\[1.25ex]
&
\bar u(p_2,\lambda_2) \big[ H^q(x,\xi,t) \, \gamma^\mu +
E^q(x,\xi,t) \, \frac{i}{2m}
\sigma^{\mu\nu} q_\nu \big] u(p_1,\lambda_1).
\end{align}
(In the frame we use, $\xi \equiv -2q^+/p^+ = 0$.)  We also have $H^g$ and $E^g$ due to gluons. From the definitions,
\ba
A(Q^2) &=& \sum_{a=g,q} \ \int_{-1}^1 x \, dx \, H^a(x,\xi,Q^2)	,
	\nonumber \\
B(Q^2) &=& \sum_{a=g,q} \ \int_{-1}^1 x \, dx \, E^a(x,\xi,Q^2)	.
\ea

For the GPDs we use the ``modified Regge model'' of Ref.~\cite{Guidal:2004nd}.  There we find,
\ba
\label{eq:valquarkgpd}
H^q(x,0,Q^2) &=& q_v(x) x^{\alpha' (1-x) Q^2}	, 	\nonumber \\
E^q(x,0,Q^2) &=& \frac{\kappa^q}{N^q} (1-x)^{\eta_q} q_v(x) x^{\alpha' (1-x) Q^2}  ,
\ea
with
\begin{align}
&\alpha' = 1.105 {\rm\ GeV}^{-2},  && \eta_u = 1.713 , 
&  							&	& \eta_d = 0.566 ,  \nonumber \\
& \kappa_u = 1.673	,		&& N_u = 1.475 ,	&	\nonumber \\
& \kappa_d = -2.033	,			&& N_d = 0.9083 .	&
\end{align}
The values of $\kappa^{u,d}$ are obtained from $\kappa_p = (2/3)\kappa_u - (1/3)\kappa_d$ and
$\kappa_n = -(1/3)\kappa_u + (2/3)\kappa_d$.

For the momentum density we need sea and gluon distributions.  The gluon GPD $H^g(x,0,Q^2)$ is discussed in~\cite{Frankfurt:2006jp}.  For applications, they give two forms, each of which is a gluon parton distribution multiplied by either a dipole form or a gaussian form for the $Q^2$ dependence.  The results from these two forms are very close to each other.  A third possibility is to use the same $Q^2$ dependence as for the valence quarks, Eq.~(\ref{eq:valquarkgpd}).  This gives a result in between the results from the two choices in~\cite{Frankfurt:2006jp}, so will be the choice we use in this paper.  We follow a similar procedure for the sea quark contributions.

All quark and gluon distributions are taken from the MRST2002 global NNLO fit~\cite{Martin:2002dr} at input scale $\mu^2 = 1$ GeV$^2$,
\begin{align}
&u_v(x) = 0.262 x^{-0.69} (1-x)^{3.50} \left( 1 + 3.83 x^{0.5} + 37.65 x \right), \nonumber \\
&d_v(x) = 0.061 x^{-0.65} (1-x)^{4.03} \left( 1 + 49.05 x^{0.5} + 8.65 x \right), \nonumber \\
&S(x) = 0.759 x^{-1.12} (1 - x)^{7.66} (1 - 1.34 x^{0.5} + 7.40 x)  ,  \nonumber \\
&g(x) = 0.669 x^{-1} (1 - x)^{3.96} (1 + 6.98 x^{0.5} - 3.63 x)        \nonumber \\
& \hskip 4 cm   - 0.23 x^{-1.27} (1 - x)^{8.7} \,.
\end{align}

With these definitions, we calculate
\be
\rho^+_{1/2}(b) = \int_0^\infty \frac{dQ}{2\pi} \, Q J_0(bQ) \, A(Q^2) \,,
\ee
where, $J_0$ is the Bessel function, $b = |\vec b_\perp|$, and $Q = |\vec q_\perp|$.

\begin{figure}[t]

\begin{center}
\vskip -24 pt
\hskip -2 mm \includegraphics[width = 7.5 cm]{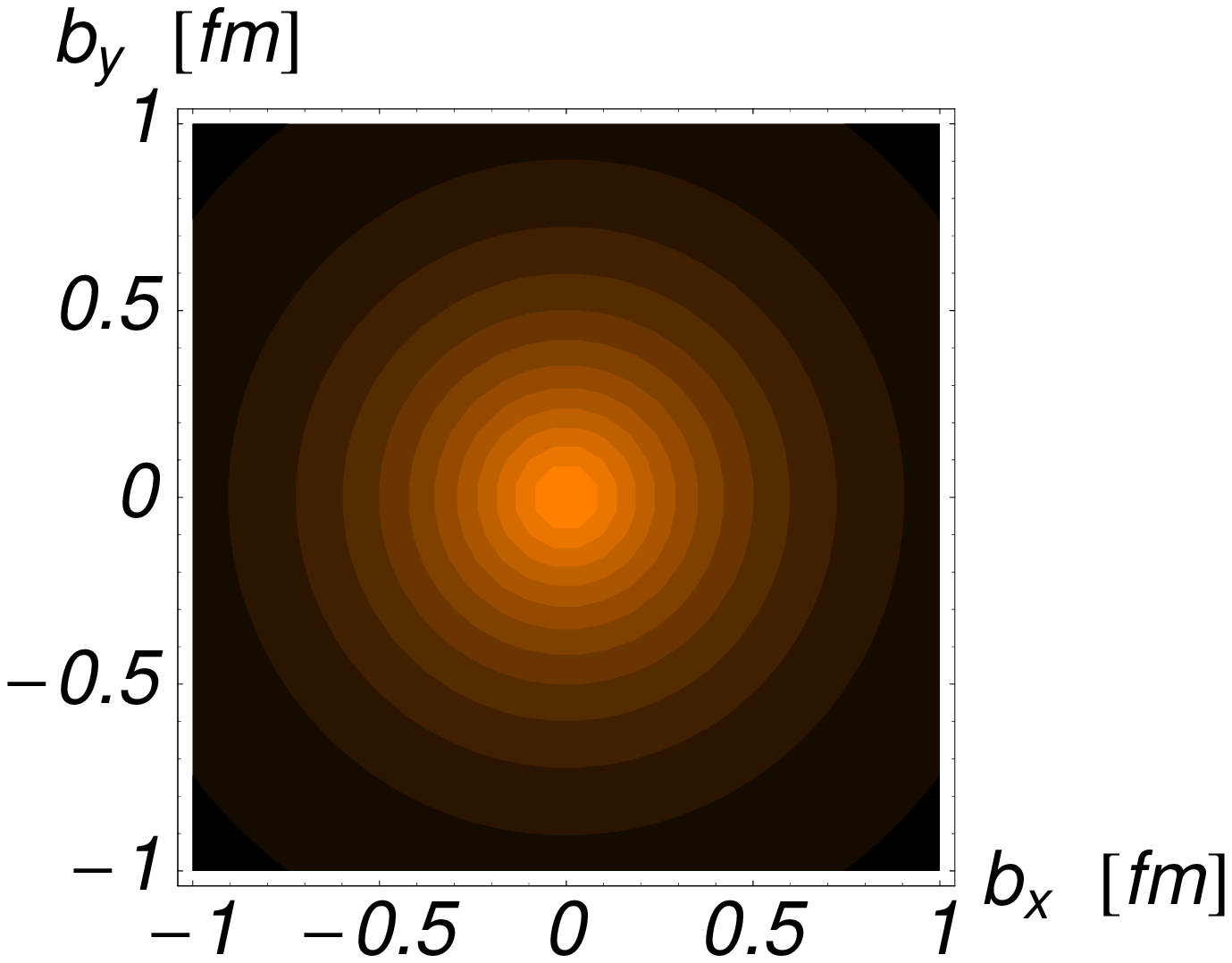}
\vskip -12 pt
\includegraphics[width = 6.5 cm]{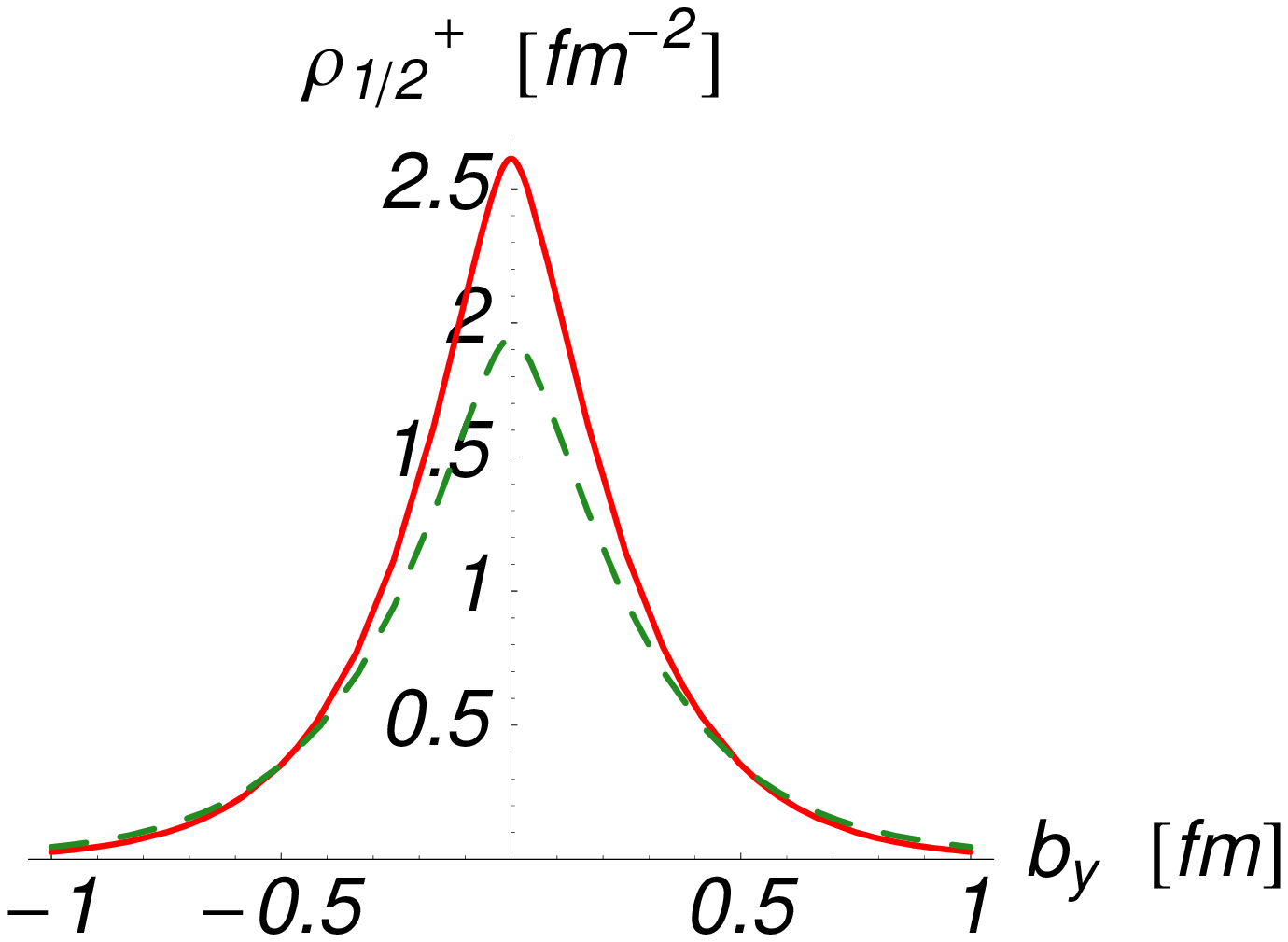}
\end{center}

\caption{Upper panel:  Momentum density (for ``$p^+$'') of the nucleon,
$\rho^+_{1/2}$, projected onto the transverse plane.  Lower panel:  The solid red line is
$\rho^+_{1/2}$ for the nucleon, along an axis through the nucleon's center. For comparison, the proton charge density is also shown, as a dashed green line.  The charge density spreads out more than the momentum density.}
\label{fig:rhoplusN}

\end{figure}

Results for $\rho^+_{1/2}(b)$ are shown in Fig.~\ref{fig:rhoplusN}, both as a contour plot and along a line through the center of the nucleon.  The result is the same for the proton and neutron, assuming isospin symmetry.  For comparison, the charge density of the proton is also shown.


The RMS radius of the $p^+$ density is $0.61$ fm, notably smaller than the Dirac radius of  $0.80$ fm (corresponding to charge radius $0.87$ fm) in this parametrization.

Nucleons can also display density shifts due to polarization.  We polarize the nucleons in the transverse plane, and denote the polarization direction by
$\hat S_\perp = \cos\phi_S\, \hat e_x + \sin\phi_S\, \hat e_y$. The state with spin projection $s_\perp = 1/2$ in this direction will be
$| p^+, \vec q_\perp/2, s_\perp \rangle$.  The impact parameter direction is denoted by $\hat b_\perp  = \cos\phi_b\, \hat e_x + \sin\phi_b\, \hat e_y$.   The $p^+$ density of the transversely polarized state is
\begin{align}
& \rho^+_{T s_\perp} (\vec b_\perp)  = \frac{1}{2 {p^+}^2} \int \frac{d^2 q_\perp}{(2\pi)^2}
	e^{-i \vec q_\perp{\cdot} \vec p_\perp}
												\\[1.2ex]
& \hskip 15mm  \times	
\langle p^+, \frac{\vec q_\perp}{2}, s_\perp |	\,
	T^{++}(0)		\,
	| p^+, - \frac{\vec q_\perp}{2}, s_\perp \rangle
												\nonumber \\ \nonumber
&= \rho^+_{1/2} (b_\perp) + \sin (\phi_b-\phi_S)
	 \int_0^\infty \frac{dQ}{2\pi} \, \frac{Q^2}{2m} J_1(bQ) \, B(Q^2) .
\end{align}

For the valence quarks, $B(0) \approx -0.02$, and $\rho^+_{T s_\perp}$ is essentially the same as $\rho^+_{1/2}$, in contrast to the corresponding electromagnetic case~\cite{Miller:2007uy,Carlson:2007xd}, and we show no plots.

%
%

{\textit {For spin-1 particles,}} the stress tensor matrix elements can be expanded in term of six Lorentz structures multiplying six gravitational form factors,
\begin{align}
& \left\langle p_2,\lambda_2 \right|  T^{\mu\nu} \left| p_1, \lambda_1 \right\rangle = \varepsilon^*_{2\alpha} \varepsilon_{1\beta}
						\nonumber \\
&	\times	\Big\{ -2 A(q^2) \eta^{\alpha\beta} p^\mu p^\nu
	- 4 \big(  A(q^2)+B(q^2)  \big)
	q^{[\alpha} \eta^{\beta](\mu} p^{\nu)}
						\nonumber \\
&	+ \frac{1}{2} C(q^2) \eta^{\alpha\beta} \left( q^\mu q^\nu - q^2 \eta^{\mu\nu} \right)
						\\
&	+ D(q^2) \left[ q^\alpha q^\beta \eta^{\mu\nu} -2 q^{(\alpha} \eta^{\beta)(\mu} q^{\nu)}
					+ q^2 \eta^{\alpha(\mu} \eta^{\nu)\beta} \right]
						\nonumber \\  \nonumber
&	+ E(q^2) \frac{q^\alpha q^\beta}{m^2} p^\mu p^\nu
	+ F(q^2) \frac{q^\alpha q^\beta}{m^2} \left( q^\mu q^\nu - q^2 \eta^{\mu\nu} \right) \Big\}\,.
\end{align}
where $a^{[\alpha}b^{\beta]}=(a^\alpha b^\beta-a^\beta b^\alpha)/2$.

The two independent helicity conserving form factors $\mathcal{T}^+_{11}$, and $\mathcal{T}^+_{00}$, defined by Eq.~(\ref{Tmatrixelement}), can be expressed in terms of the above form factors,
\ba
\mathcal{T}^+_{11}&=&A+\eta E  =  Z_2(Q^2)\,,    \\ \nonumber
\mathcal{T}^+_{00}&=&\left(\left(1+2\eta\right)A+4\eta B-2\eta D-2\eta^2 E\right)
	= Z_1(Q^2) \,,
\ea
where $\eta=Q^2/(4m^2)$.  The first equality in each line above is generic;  the second uses an AdS/QCD model worked out for rho (and other) mesons in~\cite{Abidin:2008ku, Abidin:2008hn}, where
\ba
Z_1(Q^2)&=&\frac{1}{m^2_\rho}\int \frac{dz}{z} e^{-\Phi}\mathcal{H}(Q,z) \partial_z\psi(z) \partial_z\psi(z)\,,\nonumber\\
Z_2(Q^2)&=&\int \frac{dz}{z}e^{-\Phi} \mathcal{H}(Q,z) \psi(z) \psi(z). \label{Z1Z2}
\ea
We obtain the  densities
\ba
\rho^+_1(\vec{b}_\perp)&=&\int_0^\infty \frac{dQ}{2\pi}Q J_0(bQ)Z_2(Q^2)\,,\nonumber\\
\rho^+_0(\vec{b}_\perp)&=&\int_0^\infty \frac{dQ}{2\pi}Q J_0(bQ)Z_1(Q^2).
\ea

In the hard-wall model~\cite{Erlich:2005qh}, $\Phi(z)=0$ and the $z$-integration in Eq.~(\ref{Z1Z2}) is from $0$ to $z_0$. The profile function $\mathcal{H}(Q,z)$ and wave functions $\psi(z)$ are
\ba
\mathcal{H}(Q,z)&=&\frac{1}{2}Q^2z^2\bigg(\frac{K_1(Qz_0)}{I_1(Qz_0)}I_2(Qz)+K_2(Qz)\bigg)\,,\nonumber\\
\psi_n(z)&=&\frac{\sqrt{2}}{z_0 J_1(m_n z_0)}z J_1(m_n z)  \,.
\ea
They satisfy the boundary conditions $\mathcal{H}(Q,0)=1$, $\partial_z\mathcal{H}(Q,z_0)=0$, $\psi(0)=0$, and $\partial_z \psi(z_0)=0$. The value of $z_0 = (323 {\rm MeV})^{-1}$ is fixed by the rho meson's mass, such that $J_0(m_n z_0)=0$, where the lightest mass (labeled $n=1$) corresponds to the rho meson.  See also~\cite{Grigoryan:2007vg,Brodsky:2007hb}

For the soft-wall model~\cite{Karch:2006pv}, $\Phi(z)=\kappa^2z^2$. The mass eigenvalues are given by $m^2_n=4(n+1)\kappa^2$, with now $n=0$ corresponding to the rho meson. The integration region in Eq.~(\ref{Z1Z2}) spans from 0 to infinity. Instead of boundary conditions at the upper limit, we require a normalizable wave function, $\int (dz/z)e^{-\kappa^2 z^2}\psi^2=1$, and finiteness of $\mathcal{H}$ at infinity. Using results of H. Grigoryan~\cite{GrigoryanGFF},
\begin{align}
\mathcal{H}(Q,z)&=   \Gamma(4\eta+2) \, U(4\eta, -1, z^2)
												\nonumber \\
\psi_n(z)&=\kappa^2z^2 \sqrt{\frac{2}{n+1}}L^{(1)}_n(\kappa^2z^2)  .
\end{align}
where $L^{(\alpha)}_n$ is the generalized Laguerre polynomial and $U(a,b,w)$ is the $2^{\rm nd}$ Kummer function.  

\begin{figure}[t]

\begin{center}
\includegraphics[width = 6.5 cm]{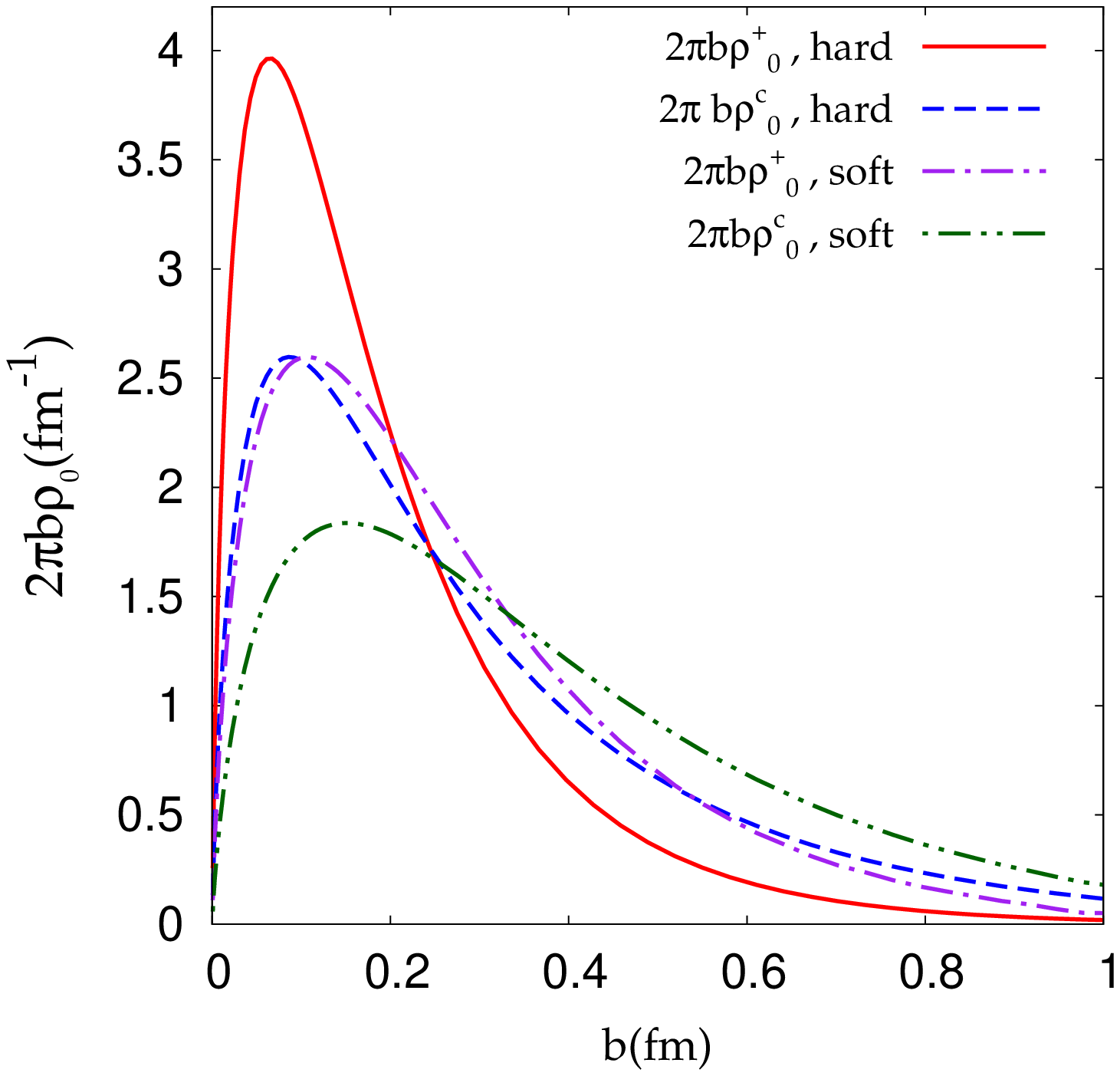}
\includegraphics[width = 6.5 cm]{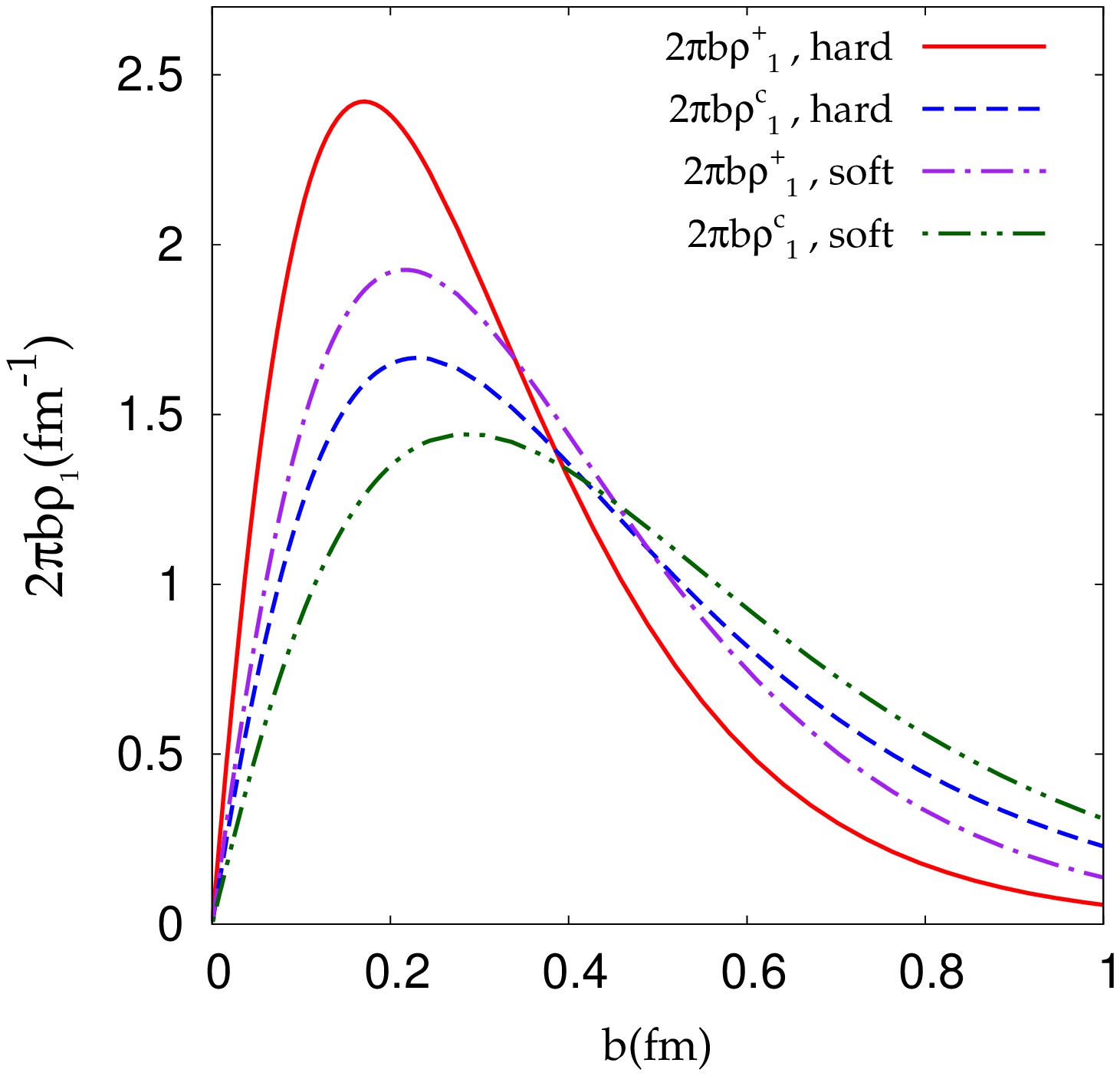}
\end{center}

\vglue -4 mm
\caption{Upper panel:  The red solid line is $2\pi b$ times $\rho^+_0(b)$, the $p^+$ density of helicity-$0$ $\rho$-mesons in the hard-wall AdS/QCD model, while the purple dash-dotted line is the corresponding result in the soft-wall model. The blue dashed line is $2\pi b$ times $\rho^c_0(b)$, the charge density of helicity-$0$ $\rho$-mesons in the hard-wall model, while the green dash-dot-dotted line is the corresponding result in the soft-wall model.  Lower panel: the same but for $\rho^+_1(b)$ and $\rho^c_1(b)$.}
\label{fig:0plots}
\end{figure}

An analytic  form can be obtained for density $\rho^+_0(\vec{b}_\perp)$,
\be
\rho^+_0(\vec{b}_\perp)=\frac{m^2_\rho}{\pi}K_0(\sqrt{2}m_\rho b). \label{softHelicity0}
\ee
The other densities can be calculated numerically. They are shown in Fig.(\ref{fig:0plots}), along with the charge density~\cite{Carlson:2008zc}.

Light-front longitudinal densities as well as charge densities, for helicity-$0$ rho-meson are logarithmically divergent at the origin for both hard-wall and soft-wall model, which is evident in (\ref{softHelicity0}) for the soft-wall model. The hard-wall model is more compact than the soft-wall model and the helicity-$0$ rho mesons are more compact than the helicity-$1$ rho mesons. Overall, the distribution of longitudinal momentum in position space is more compact than that of the charge.

The two independent helicity flip form factors are
\be
\mathcal{T}^+_{10} = \sqrt{2\eta}\left(-B+\eta E\right) \,, \qquad 
\mathcal{T}^+_{-1,1} = -\eta E.
\ee
However, both $B$ and $E$ vanish in the AdS/QCD model.

%
%

{\it In conclusion,} we have studied the distribution within extended objects of the matter that carries the component $p^+$ of the momentum, in a light front viewpoint.

  The examples we used were real nucleons, where we used semi-empirical models of the nucleon GPDs as underlying input, and spin-1 particles, where the underlying input came from AdS/QCD studies of these states.    The crucial gravitational form factors can be obtained as second moments of the GPDs.  There are conceptual similarities to the light-front relations of charge distribution in the transverse plane to Fourier transforms of the electromagnetic form factors.   Differences include using the gravitational instead of electromagnetic form factors and weighting the GPDs with $x$ instead of charge.

We presented plots that showed the $p^+$ density in the entire transverse plane.  A qualitative result is that the hadrons we study are all more compact when looking at the $p^+$ momentum density than when looking at the charge (or magnetic) density.    
We had earlier calculated ``gravitational radii'' from the slope of the gravitational form factors obtained for several species of mesons in an AdS/QCD model~\cite{Abidin:2008ku,Abidin:2008hn}.
In addition, we have learned that the phenomenon of compactness of the momentum distribution and the corresponding smaller root-mean-square radius is not limited to mesons which are studied using a purely theoretical AdS/QCD correspondence, but is also seen in nucleon distributions based on real data.

We thank  the NSF for support under grant PHY-0555600.

\end{document}